\documentclass[a4paper,12pt]{article} 
\usepackage[applemac]{inputenc}
\usepackage[british,french]{babel} 
\usepackage{setspace}
\usepackage{verbatim} 

\usepackage[hyperindex]{hyperref} 

\newcommand\bi{\begin{itemize}} 
\newcommand\ei{\end{itemize}} 
\newcommand\ben{\begin{enumerate}} 
\newcommand\een{\end{enumerate}} 
\newcommand\bd{\begin{description}} 
\newcommand\ed{\end{description}} 

\title{Who cares about physics today? \\ 
A marketing strategy for \\ the survival of fundamental science \\
 and the benefit of society 
}
\author{
{\bf Umberto Cannella}, visiting postdoc at \\
the Maryland Center for Fundamental Physics~\footnote{The opinions 
expressed in this paper are solely the author's and should not be understood 
as reflecting any official position of the hosting research institution. 
Email: \href{umberto.cannella@gmail.com}{umberto.cannella@gmail.com} } \\
University of Maryland, College Park (USA)
}

\usepackage{fancyhdr}
\usepackage{lastpage} 
\begin{document}
\selectlanguage{british} 
\maketitle

\section*{Abstract}

It would seem that the present dry economic times impose a very precise focus for science in general and physics in particular: research, possibly of applied type. 
However in doing so two basic pillars of a healthy future for science are being undermined: fundamental research and public engagement. 
The first is what makes applications possible in the first place, many times with a path from inception to implementation that is as long and indirect as poorly advertised. 
The second pillar, public engagement, is mostly regarded as a luxury: if there is good level of funding scientists can consider spending money for "public relations" otherwise this is the first thing scientists cut because it is the least necessary. 
On the contrary, public engagement in science is very much needed, at the very least because the public is either an enemy or an ally, as testified respectively by the climate change denial and the 2009 Shuttle mission that people wanted in order to service the Hubble Space Telescope for the last time.  
In this article I will make the case for why popularizing science should be a funding priority, for both nation-wide organizations and local research institutions.
I will take examples from my personal background with the hope that they will serve to enlighten a more general picture and to frame the discussion around concrete issues and practical avenues to be pursued immediately.

\section{Introduction}

It is not uncommon to hear scientists regret the poor appreciation of research by society. 
This regret regularly makes the subject of lunch chats among fellow scientists when cuts to research funding are being discussed by the government of the hosting country.
The conversation usually terminates with a recommendation that each of the fellows should subscribe some online petition letter to be sent to some agency. 
End of the story. But does it really have to be the end of the story? Is it the best the physics community can do? 
I would say no, that we can speak much louder than that, at least if we really want to have a hope for things to change; however, we have to talk to the right people and these people are most certainly not our fellow colleagues. 
One might object that cuts happen because there is no money; however the recent {\it Science and Engineering Indicators 2012}\, from the National Science Foundation (NSF) reveal a different truth: 
money for science is there, it is just that physics is not seen as a priority, let alone theoretical physics~\cite{NSFdata}. 
Despite this solid evidence, the common perception in the physics community is that the blame is to be put on the current dry economic times, which impose a very precise focus for science in general and physics in particular: research, possibly of applied type. 
Unfortunately, if these are the exclusive priorities two basic pillars of a healthy future for physics and science are being undermined: fundamental research and public engagement. 
The former is what makes applications possible in the first place, many times with a very indirect path from inception to implementation that should be stressed more often, although not in the terms of an apologetic justification. 
Reminding people that they can get their tumors cured by something called hadron-therapy, whose name is inextricably linked to the Large Hadron Collider (LHC), is not political talk, it is an integral part of the truth about fundamental research: in fact, this deep connection between nuclear physics and medicine is the first selling point for LHC presented to funding agencies, as Doctor Rolf Heuer, CERN's Director-General, told me at the end of his seminar at the University of Maryland in the Spring of 2011.

What I called the second pillar of a healthy future for physics and science, public engagement, is mostly regarded as a non-essential nice-to-have by scientists: if there is good level of funding they can consider putting some money in what is usually called "public relations"~\footnote{The reason why the term "public relations" is in quotes is because it has a sort of negative aura, as if it were a synonym for a frill and, then, not a priority.}\,, otherwise it is the first thing to be cut because it is the least necessary ingredient to advance knowledge. 
In this article I will make the case for the very opposite viewpoint: popularizing science should be a priority in the agenda of both nation-wide organizations and local research institutions. 
Compelling reasons for why public engagement is needed, today and in the future, will be given soon; for the moment I will provide just one rationale for it: the unrenounceable support from the public, most clearly exemplified by the 2009 Shuttle mission that serviced the Hubble Space Telescope for the last time. 
Previously, the telescope had been declared doomed by US President George W. Bush and NASA President Sean O'Keefe, in charge at the time; an unprecedented movement of popular opinion grew to such a large extent that the official decision had to be changed and money reallocated: Mars' exploration projects were not the priority anymore because common people had decided otherwise. 
This fact should unequivocally prove that the public is either an ally or an enemy, the last case being proved by the climate change controversy, just to quote an example.  
The public is eager to be a part of an ongoing exchange with the scientific community, at the very least because it is tax-payers' money which is being used (when it is research funded by private money I would suspect the feelings from the public to be less strong). 
Much in the same way as everyone of us wants to know from a doctor what s/he is going to do with our bodies, the public wants to know what scientists are going to do with its  money. Hence one of the needs for engaging the public. 
Other motivations that, in the body of the paper, will be tied to specific outreach initiatives are:
\bi
\item inspiring young generations and, through inception of a passion, giving them an aim, both in their careers and in their lives; when applied to societies whose composition is increasingly diversifying the inspirational element might have a crucial role in granting a qualified workforce for the future. 
\item exploring otherwise unconceivable ideas, which is the only means toward the creation of new applications and new jobs; you are not likely to become an innovator or an entrepreneur without cross-disciplinary exposure~\cite{JobsLazaridis};
\item providing people with a scientific mindset, i.e. the tools for critical thinking that would enable them, for example, to quantitatively approach the facts supporting climate change or a speech about economic growth from a presidential candidate.
\ei 

\smallskip

There is little doubt that the public has to be engaged: a part of it already wants to be, a larger part still has to be. 
In order to do that a specific language and a tailored proposal have to be adopted, just like what happens in every marketing strategy: you have to know your target public and you have to touch particular chords in specific ways.
This is successfully achieved by employing a dedicated workforce: not just anyone but someone who is gifted and inclined. 
A major point of this white paper is advocating for a truly efficient marketing strategy for science: if practicing scientists feel like directly committing to popularizing the science they produce they could go for it, otherwise they should not be forced to divert time and efforts to deal with something they might not be inclined for.
As a matter of fact "excellent researcher" and "excellent popularizer" are not necessarily synonym with each other: there is no blame on anyone in stating this, it is just coping with reality and calling things with their name. 
Carl Sagan is the prototype of an excellent popularizer, while Richard Feynman is an exception because he was also an excellent researcher. 
As much as we can welcome exceptions, we can not solely rely on these champions; rather, we should let them play their role in collaboration with  the Carl Sagan's and a dedicated body of professionals who might not be willing to practice science but are both inclined toward popularizing it and good at it. 
	This attitude solves the ambiguity according to which practicing scientists have to find the time to popularize their own science. 
	Employing a specific workforce is, instead, a more efficient strategy, both from the point of view of impact and of money needed: in fact, if you are not achieving the goals of your popularization investment, however small, aren't you wasting your money?
	Of course this does not mean that practicing scientists who are neither Feynman's nor Sagan's have no role: how else would science advance and what will be there to be popularized if there were no scientific conquests?
	Moreover research results might come in a "ready-to-be-popularized" form that is peculiar ability of practicing scientists: see for example the compelling sonification of real and simulated physical data~\cite{Sounds}.
The community of researchers and the popularization experts should be seen 
as two branches of the same network: each branch has its own specifities and skills but most of the goals are common and could be better achieved by working together and supporting each other. 

A change in paradigm is then in order: in the same way as marketing has techniques that prescind from the product, the key to a successful public engagement is to play according the rules of communication, not of science. 
	At the very least this entails that absolute completeness of information, technical nomenclature and fine details are to be banned, as the following example details unequivocally. On occasion of a public lecture about the Universe that I attended recently, a prominent scientist was talking about the cosmic background radiation, using the term "quantum fluctuations" with no explanation whatsoever. 
	Even worse, when someone from the public dared to ask, at the end of the lecture, for a more mundane explanation the speaker answered something like: "I cannot really use a different expression for it because this is the concept". 
	This is a clearest exemplification of what has been rightfully defined as "the curse of knowledge"~\cite{BookStick}: to be at the same time a master and a prisoner of the inner workings of a discipline and its technical language.
	The need of using a different type of language, or at least a more verbose and explanatory vocabulary, is motivated by the fact that listeners exist and are somewhat numerous: take, for instance, the success of popular books about theoretical physics by Stephen Hawking, Brian Greene, Lisa Randall, Lawrence Krauss, Kip Thorne, etc. 
Another interesting example in this context is the {\it Science Entertainment Exchange}, "a program of the National Academy of Sciences that connects entertainment industry professionals with top scientists and engineers to create a synergy between accurate science and engaging storylines in both film and TV programming"~\cite{SciEntEx}\,.
The whole research community should unanimously congratulate these scientists for their efforts because they contribute to introduce physics in the pop culture.
	Such a commitment by scientists has a unique role in disseminating physics and, equally importantly, in exciting a sense of fascination about it: with the first you can hope to build a scientific mindset and critical thinking, with the second you can hope in the growth of future generations of scientists; on one side you are then educating future tax-payers, voters and politicians, on the other you are recruiting future students of the so-called STEM~\footnote{Acronym for "Science, Technology, Engineering and Mathematics".} subjects. 

\smallskip 

Besides the intrinsic value of dissemination and fascination another very good reason for the scientific community to be aiming at a "cool" role in the public's eyes is private funding:
\bi
\item James Cameron, the famous movie director, has recently entered a partnership to mine asteroids~\cite{MineAsteroids}.
\item Microsoft co-founder Paul Allen has a telescope named after him, the Allen Telescope Array, thanks to his interest and economic support in the Research Institute SETI (Search for Extra-Terrestrial Intelligence)~\cite{SETIAllen}.  
\item Tesla Motors founder Elon Musk recently committed to fund part of a US museum dedicated to Nikola Tesla, apparently one of his heroes~\cite{TeslaMotors}; incidentally, I would like to mention that the initiative has collected the money needed in very few days with an enthusiastic support from the public~\cite{TeslaMuseum}. 
\item Pop singer and producer Will.I.Am has a Foundation called {\it i.am.angel}\, which, together with Discovery Education, announced a \$10 million classroom education initiative that will reach 25 million students annually, including many from underserved communities~\cite{IAmAngelFndtn}. 
{\it Focused on science, technology, engineering, arts and mathematics educational themes, the Discovery Education initiative will incorporate NASA content and space exploration themes as part of the curriculum;} 
\item The University of Delaware has received a US \$5 million gift from the {\it DuPont} Science Company~\cite{DuPont} to support the construction of the 197,000-square-foot Interdisciplinary Science and Engineering Laboratory (ISE-Lab) on the University’s Newark campus~\cite{Delaware}. 
\ei
On this basis I can't see why an entrepreneur or a philanthropist should not get fascinated by, say, particle physics and gravitational wave astronomy; I hope the future appearance of two movies will provide the fame the fields deserve~\footnote{I warned the reader about my personal bias; however, it is also to be reminded that different science disciplines might benefit one from the success of the other.}\,: {\it Interstellar}, from director Christopher Nolan and Caltech Physics Professor Kip Thorne~\cite{Interstellar}, and {\it Particle Fever} by Johns Hopkins Physics Professor David Kaplan, which tells the story behind the LHC accelerator~\cite{ParticleFever}. 
These two movies will probably do an excellent job at advocating for {\it Big Science}, a term coined by Stephen Weinberg in an account he has recently written~\cite{Weinberg}. 
In his short essay the Nobel Laureate voices his worries of a future hiatus in big physics experiments, such as the successors of the Large Hadron Collider and of the Hubble Space Telescope~\footnote{Respectively, a proposed linear accelerator and the ongoing James Webb Space Telescope.}\,. 
In my humble opinion it is very probable to witness such hiatus if the public is not heavily engaged from today. 
A precedent hiatus, cited by Weinberg, started when the US decided not to build their own big accelerator, before the LHC was even funded, due to merely political reasons. 
This reminds me of the "Hubble rebellion" I mentioned earlier on: I think that had the public shared the charm and usefulness of accelerator science, things could have gone differently. 
Remarkably, elsewhere Weinberg acknowledged his frustration in not being able to compete with space science in the eyes of the public because of the much more visual character of the latter, which makes it simpler for the public to grasp than quantum field theory. 
	
\smallskip 

	For what said so far, to guarantee the survival and health of physics and its branch of fundamental research the scientific community has to find agreement and sponsor the implementation of new ways to get to the public. 
This is even more true if one wishes to increase the number of students who choose a career in science: there is the need to expose them to successful and approachable scientists, to the science they do and to the enthusiasm they have for science~\cite{Nico1}. 
Such exposure, especially during a student’s early years in middle school, can have a profound and lasting effect in the student’s future decisions~\cite{Nico2,Nico3,Nico4}. 

\smallskip 

These dry economic times should not be taken as an excuse to postpone engaging the public in a stable relationship with the scientific community. 
On the contrary today's situation is one of now or never. 
In what follows I will offer my very concrete recipe: I will present a variety of public outreach initiatives that stem from the communications needs of a Physics Department, take advantage of cross-disciplinary collaborations with the whole University and represent an investment for the growth of the latter.
It goes without saying that the variety of initiatives proposed in this article is synonym with variety of target public: young boys and girls, general public, women and minorities can all be found as typical audience in one or more activity. 

	The initiatives I propose originate from a few personal elements: my background in theoretical physics, my exploration of the outreach world and my life and work experiences in Italy, Switzerland and the US. 
	The specific examples help framing the discussion around very practical avenues for immediate implementation; on the other hand the situation they serve to counteract is common not only to the US but to most of the G20 countries.  
	
\smallskip 

Before going to the crux of the proposal I will briefly summarize the main motivations and aims.

\subsection{Motivations: summary}

\begin{description}

\item There is a basis of public curiosity that should be addressed as the starting point of a conversation. 
With the announcement of the discovery of a Higgs-like particle many people want to know what the {\it God particle} is~\cite{HiggsPop}. 
Around the time the accelerator was to turned on LHC and CERN were feared to both create black holes~\cite{LHClawsuit} and to be the place where anti-matter could be stolen in order to make bombs, as in the bestseller book {\it Angels and Demons} from Dan Brown and the subsequent movie inspired by it. 
Notwithstanding scientific imprecisions, CERN took an active role in parallel to this negative advertising campaign: the strategy implemented by CERN is an excellent example of science communications along the lines advised in this proposal~\cite{AngelsDemons}.

\item Fighting lack of interest before lack of scientific literacy at all levels: workforce, politicians and taxpayers = those who give money to government agencies.

\item Public is powerful: remember when Hubble Space Telescope had been declared doomed? The public can be an ally. 

\item If scientists do not talk to the public, someone outside science does: do we like the consequences?  

\item Physics departments all around the US are threatened by closure ... maybe there are no more students willing to take increasing loans and study subjects whose jobs don’t seem to be there. 

\item No ideas today = no applications tomorrow (technology, jobs). 

\item Today US are no.1 in science for results and inspiration but what about tomorrow?  

\item On May 17, 2012 {\it the U.S. Census Bureau released a set of estimates showing that 50.4 percent of our nation's population younger than age 1 were minorities as of July 1, 2011. This is up from 49.5 percent from the 2010 Census taken April 1, 2010. 
A minority is anyone who is not single-race white and not Hispanic.}~\cite{Census2010}

This poses a challenge for both education policy and the future job market: who will study physics tomorrow if today already it is not very popular?
In fact, {\it while 30\% of white Americans has a degree, the percentage drops to 18\% in the case of African Americans and to 13\% for US citizens of Hispanic origin.}~\cite{Census2010} 

\end{description}

\subsection{Advantages and aims: summary}

	\begin{description}	  
	\item Successfully addressing {\it all} of the previous issues.  
	\item More research funds: having a project for a consistent public outreach program is required by NSF when you apply for a Physics Frontier Center. It really is not just a matter of public affairs. 
	With sequestration cuts around the corner fighting for survival is not just a scary title for a proposal such as this one~\cite{sequestration}. 
	\item Building inter-disciplinary collaborations among University departments to encourage new ways of doing science communications. Fostering these collaborations through students involvement is key to teach critical thinking and entrepreneurship, provide job experience and  diversify the curriculum.
	Science communications is an investment for a University. 
	\item More public engagement, in general and from under-represented communities.  Getting minorities engaged also means integrating them in the society. 
	\item Giving science communications both a marketing character and a lobbying aspect: these two features are equally crucial to make the most of individual laudable efforts and to reach the critical mass needed to influence public opinion and public policy.  
	\end{description}

\section{How}

\subsection{First things first: being visible on internet}

\subsubsection*{Social Media}

In the Introduction I pointed at the crucial role of proactively going toward the public: in this context a privileged place where having a conversation with people is internet and social media. 
Not surprisingly research organizations such as NASA and CERN are active on various social media platforms, including but not limited to the more notorious Facebook and Twitter.
The Laser Interferometer for Gravitational Wave Observatory (LIGO) and the Perimeter Institute for Theoretical Physics in Canada have quite a following too: a clear sign of the fact that the public is indeed curious and eager to know even about theoretical and fundamental physics, although through ways which are unconventional from the purely academic dissemination means.

\subsubsection*{A scientific Web Blog and a scientific Wikipedia project} 

To keep alive the conversation with the public there might be the need for a blog (=web blog). A blog where there is more than one author~\cite{SharedBlog} could be the best choice for a variety of reasons: it would serve the purpose of covering diverse physical domains, it would reflect the image of a collective Physics Department effort, it would have more impact than individual efforts, it would not ask commitment from a single person. 

\smallskip 

In a similar collaborative spirit, a Physics Department could participate in a new Wikipedia project, which aims to make science research accessible by offering plain-language summaries of key papers: such is the aim of the planned {\it 21st Floor Wiki Project}~\cite{WikiProject}.

\subsubsection*{Various types of videos} 
\label{sec:Videos}
On the internet videos go viral, more than anything else, probably because of their blending of words and action. 
An initiative that would have a strong impact on how physics research (especially theoretical) is perceived by society is something called FameLab~\cite{FameLab}, a sort of "American Idol" for scientists. It consists in a contest based on science communication where participants are young researchers, such as grad students and postdocs, and have three minutes to present an item of their research with nothing but props (no slides!). 
The example I like to quote the most is one of past winners' presentations~\cite{FameLabWinner}, which is particularly suited to theoretical physics.
Before reaching the nation-wide scale of an inter-university contest, young researchers at a Physics Department could record a three-minute video in front of their laptops' webcam~\footnote{I have uploaded my personal try at it on the Facebook page of the Maryland Center for Fundamental Physics~\cite{MyFameLab}\,.}\,. 
This exercise is a necessary intermediate step that would let young researchers practice at two major communications skills: talking in front of a camera and getting the message across in a limited amount of time.
Possible uses of these videos are: a dedicated tab on the Physics Department website, the University Youtube channel, the integration into open days at the University in the form of a show (similar to the finals of the FameLab contest). 

\smallskip

Another set of videos could feature interviews of senior faculty members, who were personally involved in the inception of research fields and concepts that now are part of mainstream. This is a legacy we should make sure to preserve as scientists and it offers the possibility of presenting laymen with the characters behind physics investigations, their faces and emotions.
Moreover, it constitutes an example of the concept "everyone can participate in outreach", even if not completely at ease with a personal use of internet and social media. 
Here at the Maryland Center for Fundamental Physics I am looking forward to interviewing at least three senior faculty: 
	\bi
	\item Professor Oscar Wallace Greenberg, who participated in the development of the ideas that preceded the introduction of quarks into the world of particle physics and witnessed the pioneering efforts towards the detection of gravitational waves by John Weber, a University of Maryland Professor in the 1960's. 
	\item Emeritus Professor Charles Misner, who has heavily contributed in shaping General Relativity's investigations and teaching. 
	\item Professor Rabinda Mohapatra, whose craft involves neutrino physics and Supersimmetry as examples of physics beyond the Standard Model. 
	\ei 
Personally, it is not enough for me to meet these professors in the corridor: I would like to collect for the public their first-hand accounts of physics as a work-in-progress, possibly with some tales (for instance, Professor Charles Misner knew Einstein personally).

\smallskip
Other examples of video initiatives include the following. 
MIT has recently put forth a video initiative, mainly about experiments, with target audience  represented by students from kindergarden to 12-th grade~\cite{MIT}. 
A more technical case pertains to the experimental branches of physics and comes from submitting one's research works to the Journal of Visualized Experiments, JoVE~\cite{JoVE}, which is a PubMed-indexed video journal. 

\smallskip

It is crucial that the aforementioned channels of communications with the public are properly displayed on the Physics Department's website and possibly even on the University's website itself: this attitude gives a sense of how the University values communicating to the public in order to engage it in a conversation.

\subsection{Building inter-disciplinary collaborations among University departments to encourage new ways of doing science communications} 
\label{sec:departments}

The closest public for science communications activities is definitely the University community:  
the Physics Department could start a program of modern and engaging seminars about its cutting edge research, explicitly targeting its neighboring audience. 
	In fact, in order to extend the reach and breadth of its science communications activities I suggest that the Physics Department collaborates regularly with other University departments. Examples will be given in what follows to point at ways in which contamination and cross-pollination among fields can diversify the scientific message in order to successfully bring this message to a larger audience. 
		Fostering inter-disciplinary collaborations among departments through students involvement is a key to teach critical thinking, inspire entrepreneurship, provide job experience and diversify the curriculum: these ingredients are fundamental for an outstanding curriculum and to be successful (chameleonic) in the job market. 
		In this context it is probably useful to mention that a background in physical sciences is among the skills required of professionals in job sectors that people would not immediately associate with physics, such as a medical doctor~\cite{MedicalDoc} or a designer of realistic special effects in movies and video-games~\cite{VideoGames}. 

\smallskip

Most of the collaborative initiatives that will be presented cost either a very limited amount of money or even nothing more than the budget already in place for a University, its departments and/or its scientific research groups: it is more a matter of capitalizing existing assets at a University in a concerted way and taking full advantage of them by means of new synergetic collaborations. 
In this context, science communication represents a basis for systematized cooperation and ramified networking that can give more than the sum of the individual components; in other words, science communication is an investment for a University.

\subsubsection{Collaborating with the Performing Arts Department} 
\label{sec:Arts}

\subsubsection*{Dance performances}

In the year 2010, at the University of Maryland, the Physics Department collaborated with 
the Performing Arts Department~\cite{UMDArts} to create the dance performance {\it The Matter of Origins}~\cite{Origins}: it is this type of unconventional out-of-the-classroom activities that best captures the interest of the non-expert public. 
The aforementioned program of seminars, where the Physics Department invites faculty and students from other Departments, seems to be the ideal first step toward a more stable collaboration with a workforce that has peculiar expertise in the field of communicating and expressing ideas. 
The University Schools of Theatre, Dance and Performance Studies are likely to have artists interested in designing a course of theirs around one or more scientific themes: being part of a course held at the University, realizing such a project does not cost money outside the annual budget for the departments involved. 
At the University of Maryland I have recently participated in one such project: in collaboration with Professor Cole Miller of the Astronomy Department and Adriane Fang, instructor at the School of Dance, we built the dance show {\it Gravity}~\cite{UMDGravity}, inspired by  gravitational-wave astronomy. 
In the near future other performances of this type will be displayed at Universities in the US~\cite{YunesCamp}: very interestingly, for one of them, {\it Celebrating Einstein}\,, the materials will be compiled so that other venues can host the event. 

The dialogue between science and performing arts is also taking place as a sort of grassroots effort called {\it Dance your PhD}~\cite{DancePhD}, a very smart idea for a contest where you have to explain your PhD thesis by means of a dance show. 
As an example, read how the 2011 participant Elisa House describes her PhD dance project "Cosmological Simulations of Galactic Disc Assembly"~\cite{DancePhDGalaxies}: 
{\it Our dance starts with a bang, a Big Bang. The dancers expand homogeneously like gas and the mysterious dark matter. This dance is mirrored throughout the Universe and reflects the lack of a centre. As hard as they try to ride the expansion, the dancers cannot escape as the pull of gravity forces them back into the clutches of the imperious dark matter, causing them to collapse.
Filamentary structures start to form as the dark matter takes control over the fate of the dancers. Using gravity as its lasso the dark matter pulls the dancers ever closer. Chaotic mergers, collisions and interactions of the dancers build the galaxy. These mergers become less frequent as the Universe ages allowing for more ordered motions. Satellite dancers fall into the forming galaxy late and are stripped bare.
Emerging gloriously, the galaxy swirls and spins her beautiful disc while the remaining satellites dance to her tune.}

I believe that this type of language and visual handle should become widely used by scientists when they want the public to get interested and excited about something: this is a necessary prerequisite to have the public be willing to know more about science.

\subsubsection*{Artistic projects} 
\label{sec:Art} 

The blending of art and science is something I got to know at the time of my doctoral studies at the University of Geneva: on many an occasion artists from outside the University would propose how they see and relate to the physical sciences~\cite{UniGeArt}. 
Still in Geneva, CERN has recently put forth a stable collaboration with an artist in residence~\cite{CERNArt}. 
This type of initiatives is undoubtedly valuable for public appreciation in a way that is difficult to match. 

In the spirit of the ideas presented in this article I rather propose that the artists come from the University pool of faculty/students. A distinct scope of their works could be the one of engaging and inspiring under-represented minorities, for example by drawing murals to contribute to the re-qualification of public spaces. 
In the recent past NASA Goddard has collaborated with the artists behind the {\it Big Hands} project~\cite{BigHands} to realize murals with a space-science theme.  
The University artists could probably be inspired by hearing about the research activities carried on at the Physics Department, again on occasion of the public lectures I have mentioned more than once previously.

\subsubsection{Collaborating with the Computer Science Department} 

\subsubsection*{Web design}
This department is one of the most representative in terms of providing hands-on job experience for the University students during their studies. 
Because of the necessity of the Physics Department of having a strong presence on the internet, the availability of in-house experts in web design is an invaluable asset of the University. 
Graduate students from the Computer Science Department might be paid Summer through internships to refresh the pages of individual research groups or conceive ideas for new websites specifically devoted to outreach initiatives. 

\subsubsection*{Video-games and Apps for mobile devices}
Many universities already offer courses that teach how to design video-games and Apps for mobile devices, which represent a fast growing market. 
Possible themes include scientific subjects, as in the cases of {\it LHsee} mobile App~\cite{LHsee} or the game {\it Angry Birds in Space}~\cite{AngryBirdsSpace}, which has been realized in partnership with NASA and has obtained a planetary success. 
Let me also remind here that a background in physical sciences is among the skills required of a professional designer of realistic special effects in movies and video-games~\cite{VideoGames}.

\subsubsection{Collaborating with the Journalism and Literature Departments} 

This department in particular can bring the expertise required to target different audiences with 	specific languages: kids, youth, adults. 
The idea of a partnership between Sciences and a Journalism Department, in the context of a physics outreach program, has been recently explored by Caltech in collaboration with K. C. Cole, Professor of Journalism at Southern California University~\cite{KCCole}: 
the goal is to systematize a two-way dialogue between scientists and writers so that on one side scientists develop a taste for clear and rich ways of explaining things and on the other side writers are not scared by lack of accuracy when treating scientific subjects. 

Master programs in scientific journalism are in place, for example, at MIT~\cite{MITjourn} and Johns Hopkins University~\cite{Hopkinsjourn}. 
Even outside a formalized program of this type, a stable dialogue between the Physics Department and the Journalism Department could be built by means of the public lectures cycle that is a sort of connecting thread in the collaborations I propose among University departments.

\subsubsection*{Comics} \label{sec:comics}
A targeted tool to engage young kids is represented by comics. 
A reference example is the very nice project put forward by the American Physical Society~\cite{APScomics}. 
A possible way for a University to collaborate with APS is to translate existing projects from English to Spanish and Portuguese, the languages of two big communities which are under-represented in science.

NASA Goddard has a collaboration with a writer of Manga, the Japanese take on comics~\cite{Hayanon}. 
The worldwide market of manga is characterized by very interesting figures: 
Japan, \$5.5 billion in 2009; U.S. and Canada, \$175 million in 2008; 
Europe and the Middle East, \$250 million in 2012 (source Wikipedia).
These values would justify, for example, the commissioning of a Manga issue (starting with Japanese and English) about gravitational-wave astronomy, given that Japan is building its own interferometer~\cite{KAGRA}, one of a worldwide network that also counts the US, Europe and (in the near future) India.  

\smallskip 

The {\it PhD Comics} project~\cite{PhDComics} has recently put forth a new initiative~\cite{PhDComicsTV}: have your PhD thesis illustrated and animated by Jorge Cham, the proponent of these very smart and successful ideas. 
The contest ended on August 24 (2012) with two hundred submissions and thousands of voters~\footnote{One physicist made it to the podium, Michael Marks, from the University of Bonn, with {\it Theoretical astrophysics and binary stars}\,.}\,. 
Together with the {\it FameLab} initiative of Sec.~\ref{sec:Videos}\,, participating to these contests should become a pride for research institutions in order to maintain a regular commitment to public engagement and in order to diversify the training of their workforce, along the lines mentioned in Sec.~\ref{sec:OutCenter}. 
 
\smallskip

To conclude this section I would like to quote a couple of initiatives that are editorial and artistic at the same time. 
The first is a pop-up book about the ATLAS experiment at LHC, commissioned by the ATLAS collaboration itself~\cite{PopUpBook}; 
the second is {\it Tactile Astronomy}~\cite{TactileAstronomy} of the Space Telescope Science Institute, where astronomical images taken with Hubble are adapted to be enjoyable by people with reduced visual ability. 
It would be nice to replicate both projects in other scientific contexts with input/participation of the workforce available at a University.

\subsubsection{Collaborating with the School of Languages} 

\subsubsection*{Engaging under-represented communities, part I} 

Most of the initiatives presented so far have been proposed as collaborations among the student/faculty workforce of different departments. 
The starting point for these collaborations has been indicated in a cycle of public lectures organized by the Physics Department. 
The public of these informal seminars does not have to be limited to the student/faculty body, quite the contrary: the seminars could offer a concrete possibility of opening the doors of the Ivory Tower, especially to under-represented communities. 
Moreover, with the diverse international workforce making up scientific departments, the seminars offered could turn out to be numerous. 

\smallskip 

The University of Maryland School of Languages, Literatures and Cultures has tutors and programs for such languages as Spanish and Portuguese: it would seem a direct link to collaborate with them in order to engage these linguistic communities in the University neighborhood. 
An immediate starting point is represented by translating content that is already available from the English language, for example comics, as described in Sec.~\ref{sec:comics}. 

Besides, NASA is quite active in addressing under-represented minorities, such as the Hispanic community~\cite{NASAes}: the Agency might find a helpful partner in a University with a Language Department interested in cross-disciplinary studies, another plus for students with one major in a scientific domain.

\subsubsection{Collaborating with the School of Education} 

\subsubsection*{Shaping the national workforce of physics teachers}

All over the US, physics teachers are needed who own an actual degree in physics instead of other sciences; many a program is currently being implemented to fill this lack in the job market~\cite{AMNH}. 

Recently the University of Maryland has joined the {\it UTeach} initiative~\cite{UTeach}, a program meant to increase STEM majors with teacher certification. 
So far the program has produced over seven hundred graduates, most of whom are successful STEM K-12 teachers;  
{\it UTeach has been lauded as a national model for STEM teacher preparation by several state governors, Presidents Obama and G. W. Bush, and in the National Academy report "Rising Above the Gathering Storm"}.
In such a context, the Physics Department could widen (or target) its public seminars as teacher workshops/refresher courses: in fact, the last decades of developments in physics rarely make it to the school physics courses. 

\subsubsection*{Engaging under-represented communities, part II} 

To conclude this section I want to present an example that does a great job at synthesizing the main philosophy of this article: going toward the listeners' needs in order to successfully achieve awareness and appreciation of science. 
Dr. Christopher Emdin is an african-american Professor of Science Education at Columbia University, whose scholarly interests include: issues of race, class, and diversity in urban science classrooms; the use of new theoretical frameworks to transform science education; urban science education reform~\cite{hiphopprof_col}. 
He had a very brilliant idea: to use of hip hop music in order to teach science to black and Latino kids (who make up 70\% of the city’s rolls, according to New York’s Independent Budget Office). Not only he has written the book about it, {\it Urban Science Education for the Hip-Hop Generation}~\cite{hiphopprof_pers} but he has also partnered with Neil Degrasse Tyson and rap singer GZA~\cite{hiphop}:
{\it The project is about more than trying to teach the students science. It’s also about relating to the urban youth’s interests and keeping them engaged.} 

\smallskip 

On a related note, I would like to mention my personal exploration of the matter: the days following the announcement at CERN of the discovery of a Higgs-like particle I was inspired to write an {\it Ode to the Higgs}~\cite{odehiggs}.

\subsubsection{Collaborating with Students Associations and other projects on campus}

Many Universities can pride themselves of a radio, a newspaper and/or a tv: these channels of communication should be among the regular partners of the Science Departments. 

The same holds true for Students Associations similar to the following one.
Here at the University of Maryland I have recently started interfacing with the {\it UMD Society of Inquiry}, whose main goal is {\it to provide verifiably accurate information on specific subjects of interest to the University of Maryland community}. 
It would be invaluable to involve similar associations in a series of cross-departmental seminars about, for example, scientific matters that also have an ethical, religious and/or philosophical component.

\subsection{A demonstrations laboratory as an Outreach Center}
\label{sec:OutCenter}

It is not uncommon for a Physics Department to have a physics demonstrations program: this asset could become a flagship of the scientific communication efforts once upgraded with the creation of a dedicated space, ideally like the aforementioned one at the University of Delaware~\cite{Delaware}.
For the sake of concreteness I am going to make the case for this initiative by using a reference project from my direct personal experience: the demonstrations lab of the Physics Department at the University of Geneva, Switzerland (whose cost was an order of magnitude smaller than~\cite{Delaware}).
The project, called {\it Physiscope}~\cite{Physiscope}, was created a few years ago to counteract the loss of interest about physics in both students and adults; in order to be modern and attractive it has been rightfully chosen to build the lab in a stylish technological way that pleases the eye. 
Besides the educational component for school pupils, the project also has the goal to advertise the University commitment to research and the achievements of its scientific community. 
 
There are tree main features of a demonstrations laboratory that make it stand aside from the other  initiatives I have proposed so far. 
\ben 
\item A dedicated lab space and its activities serve to establish a direct contact and a lasting relationship with the general public; it is to follow-up on the Open Days major universities organize annually and to address the interest behind the Open Days on a more regular basis. 
Moreover, if the Ivory Tower opens its doors, it becomes a better known and more attractive place, whose usefulness and proximity to the public are shared concepts. 
\item Schools are both an audience and a partner: on one side kids  attend thematic sessions, on the other the content and the implementation of the sessions is tuned with the input of a few representative teachers. 
In view of the implementation of the {\it Next Generation Science Standards}~\cite{Standards} this would seem a very desirable feature. 

\item The staff running the presentations of the demonstrations lab is composed of grad students and postdocs, as part of their Teaching Assistants duties: on occasion of brainstorming and decision meetings this assures a varied input and a multi-disciplinary viewpoint. 
Furthermore, by training young researchers as communicators, this unconventional teaching assistantship provides them with a first-hand appreciation of clarity of explanation for a diverse audience, which will reward them when teaching, writing papers and grants. 
In other words, an outreach center turns out to provide professional development for young researchers, making them better scientists and endowing them with skills that are useful in jobs outside academia too.
\een

\subsection{Large visibility initiatives}
\label{sec:HIvisib}

In parallel to the previous initiatives I would like to suggest the pursuit of large visibility initiatives, that can aim at a national or even international type of impact. 
This is the case of employing celebrities as public advocates for sciences to attract and catalyze the interest of those who are not that much into science. 
A notable example is offered by pop music artist and producer {\it Will.I.Am}\,, who's promoting NASA and research in robotics because he has embraced the cause of improving STEM education: he uses words appropriate for a navigated expert in science advocacy~\cite{IAmAngelFndtn,Will.I.Am}. 
The cost for hiring a celebrity for this kind of job is not as high as one might think: I personally know that NASA had celebrities come for free or minimum wage, thanks to the sense of social mission that the iconic characters knew they were fulfilling. 
It is a matter of finding the celebrities that would grant the most in terms of visibility and this should be enough to justify the cost. 
In this context I propose that a large University, better yet a national coalition such as the American Physical Society or the American Association for the Advancement of Sciences, works with public figures around an initiative that takes full advantage of the notoriety that the discovery of a Higgs-like particle has had recently. 
What I envision is a {\it Higgs Day}, an event that starts with a public lecture about the discovery and goes on with a panel discussion that elucidates the ties of the discovery itself with respect to science, technology, society and politics. 
The panel could be composed of the following people: 
\bi 
\item two scientists whose expertise lies in Higgs research, one for the theoretical aspects one for the experimental ones; 
\item a scientist whose expertise lies in a different scientific domain; 
\item an expert in science policy; 
\item an expert in science education; 
\item a representative of the political scene; 
\item two representatives of the University body of students, one from a scientific discipline and  one from a non-scientific discipline;
\item a public figure to whom the young and laymen audience can relate. 
\ei 
For the last of these categories I am going to make a suggestion: the most iconic candidate I can think of is Jim Parsons, the actor who plays the role of theoretical physicist Doctor Sheldon Cooper in the popular tv series {\it The Big Bang Theory}~\cite{BBT}. 
I hope the reader does not get me wrong here: I do not aim at making science being taken any less seriously than it deserves; the goal of these initiatives is always to talk about science, even more so when they are large public initiatives. What I am trying to stress is that if you want people to listen to scientific content you have to attract their attention and present something they can relate to. 
	There is a part of the public who is prone to go toward science content in a natural way. However there is a much larger part of the public who considers this content uninteresting or useless because of a predetermined set of opinions. 
When it comes to theoretical physics I surprisingly found that the lack of any interest in it and the idea that most of it is a useless speculation are not a prejudice of un-educated people: I could hear it from professionals educated in both non-scientific and scientific disciplines (and even inside physics to tell the truth). 

Incidentally let me remind here that employing famous public advocates has been successfully undertaken by NASA since the early 1970's popularity of the tv series Star Trek and of its iconic African-american character Nichelle Nicols, a.k.a. Lieutenant Uhura. 
By means of this "public relations" effort NASA has successfully engaged under-represented social and ethnic groups. 
Let me also stress that disseminating the scientific content which is the object of the individual initiative is just one of the aims: a larger goal is to help build a scientific mindset and a critical attitude; these skills are at the basis of a functional society which is able to discern scientific matters in general and, for example, understand investments policies of presidential candidates.

\section{Final Considerations}

It is often said that the mayor stumbling block for doing some or more science outreach is money because of issues such as the funding crisis; however, this statement is too generic to be quantitative or complete: in fact, as for every project or investment, the discussion should be based on context and impact; what is the price tag for changing an unsatisfactory and dangerous {\it status quo}? does not a project that allows growth and defends the future of the field weigh more than anything else?
In order to frame otherwise generic objections about money I aimed at concreteness in this proposal: initiatives are at hand that cost either a very limited amount of money or even nothing more than the budget already in place for a University, its departments and/or its scientific research groups; it is more a matter of capitalizing existing assets at a University in a concerted way and taking full advantage of these assets by means of new synergetic collaborations. 
Let me take the example of the initiative I described in Sec.~\ref{sec:Arts}\,, where I proposed a collaboration between the Physics Department and the Performing Arts Department aimed at conceiving a performance that uses non-verbal languages to communicate science. 
By direct experience I can affirm that this type of fun and rewarding interaction for a physics postdoc takes around three hours per week for one term: assuming a postdoc is payed US \$45,000 per year, to work 8 hours per day, 5 days a week, the cost of the interaction with the Performing Arts Department amounts to \$843.75~\footnote{Assuming instead a higher number of working hours per day the cost, of course, goes down; even doubling these amounts because of overheads I would still support the initiative given its character and impact.}\,; 
couldn't this money be spent in compliance with what the NSF calls {\it Broader Impact}\,? wouldn't the filmed record of the final performance be a clearest and cleanest means to introduce the general public to a group's physics research when proudly hosted on the group's website?

The cross-disciplinary initiatives that constitute the core of this proposal all share the trait of 
systematized cooperation and ramified networking at a University that can give much more than the sum of the individual components; in other words, communication is beneficial for science but is also an investment for a University, in terms of both its publicity and the training of its workforce. 
A notable example in this sense is the one of Sec.~\ref{sec:OutCenter}\,, where I discussed how an Outreach Center at a Physics Department is also a means to provide young/early career researchers with professional development, helping them become better scientists and endowing them with skills that are useful in jobs outside academia too (a very desirable feature given the clog of the hiring system in academia).

\smallskip

There are instances in which an Outreach Center extends its scope beyond physics demonstrations; this is the case of the initiative put forth at the Perimeter Institute for Theoretical Physics in Canada~\cite{PIoutreach}: here the material produced by a locally employed workforce reach an audience outside the geographical boundaries of the Institute. 
In fact, this material could be thought of as "literature" in the field: it is of high quality, it is public and it is available. 
On this basis it would be sensible to adopt for outreach the same scheme by which practicing scientists operate to exchange results about their research: publications on bulletins/journals, collaborations.  
Research institutions that can not count on the same resources of a privately funded institution could adopt educational material produced externally and be "just" users of these products: this very efficient attitude could allow professional quality outreach to be delivered even in absence of the people and the money needed for a professional outreach staff. 

\smallskip

Besides being multi-disciplinary and cross-departmental the initiatives I proposed share 
another common theme: they represent unconventional out-of-the-classroom learning for their target public; blending scientific content with non-scientific languages and expression forms is a founding feature of an efficient marketing strategy for sciences.
I like to refer to this attitude as "invasive outreach"; elsewhere, a more audacious term, that I equally enjoy, has been coined: {\it Outrageous Outreach}\,~\cite{Outrageous}. 
As detailed in the Introduction this is motivated by the fact that in the same way as marketing has techniques which prescind from the product, the key to a successful public engagement is to play according the rules of communication, not of science. 

\smallskip

In the context of informing the public about science and engaging it in a conversation  Universities and Research Institutions can play a privileged role through employing a dedicated workforce.
I just had the chance to enjoy a seminar from Professor Georgi Dvali, who was visiting the Physics Department at the University of Maryland for the day. 
He shared his most recent ideas about black holes being inherently quantum objects, even those of macroscopic size~\cite{Dvali:2012}: if confirmed this newly suggested feature would make black holes richer than thought before and even more interesting in the eyes of the public. 
Attending the seminar I could realize once more the importance of having science popularizers work where research happens and is shared among scientists, in much the same way as news reporters go where important events are taking place. 
According to this principle an Outreach Center at a University can bring science to the public in a  unique way. 
A very notable instance along these lines is the category of initiatives where actual research data is shared with students and teachers, for example: {\it QuarkNet}~\cite{QuarkNet} for particle physics, {\it NITARP}~\cite{NITARP} for astronomy and the {\it Interactions in Understanding The Universe} virtual laboratories~\cite{eLabs}\,.
On the basis of these examples the role of a University/Research Institution in bringing science to the public is not in competition with that of a museum but complementary to it. 
Moreover, as stressed in Sec.~\ref{sec:OutCenter}\,, this attitude benefits a University because if the Ivory Tower opens its doors, it becomes a better known and more attractive place, whose usefulness and proximity to the public are shared concepts.

\smallskip

It goes without saying that the variety of initiatives proposed in this article is synonym with variety of target public: young boys and girls, general public, women and minorities can all be found as typical audience in one or more activity. 

\smallskip

In conclusion, upon recognition of the urgency of properly engaging the public and the risks of not addressing this issue the entire scientific community should strongly and unanimously manifest its consensus toward the need of science ambassadors and reinvigorate advocacy in their favor. 
This means acknowledging the need for both the job that the ambassadors have to do, hopefully in collaboration with the scientists themselves, and a larger task force of them, with increased scope and possibly means. 
In other words, there is a concrete need to give science communications both a marketing character and a lobbying aspect: these two features are equally crucial to make the most of individual laudable efforts and to reach a critical mass.  
The strategy and goals of the science ambassadors' mission should be shared and endorsed at the national level and developed along very concrete and immediate lines of intervention.  
Prototypical examples of these lines of intervention and the reasons behind them are detailed in this proposal; their character is specific for the sake of concreteness: this does not spoil the urgency of acknowledging the challenge at stake and take immediate action toward it. 
Not assuming a clear role about these issues, either an active participatory one or a supportive though unengaged one, is a luxury that the scientific community cannot afford, not today, not anymore, especially with sequestration cuts around the corner.

\section*{Acknowledgments}

I would like to thank the following people for useful inputs and/or stimulating discussions:
\bi
\item Doctor Raman Sundrum, Professor at the Maryland Center for Fundamental Physics at the University of Maryland; 
\item Doctor J. Randy McGinnis, Professor of Science Education at the Department of Teaching, Learning, Policy and Leadership at the University of Maryland; 
\item the Education and Public outreach staff of the Astrophysics Section at NASA Goddard Space Flight Center, in particular Sara Mitchell;
\item Doctor Michelle Thaller, Assistant Director for Science Communication and Higher Education, Sciences and Exploration Directorate at NASA Goddard Space Flight Center; 
\item Doctor Nico Yunes, Assistant Professor of Physics at Montana State University;
\item Jessica Santascoy, Astronomy and Social Media Coordinator for the NASA Night Sky Network, managed by the Astronomical Society of the Pacific; 
\item Doctor Olivier Gaumer, responsible of the scientific content and the communication for the {\it Physiscope}~\cite{Physiscope}, the demonstrations laboratory of the University of Geneva, Switzerland. 
\ei


				\end{document}